\documentclass[preprint,12pt]{elsarticle}
\usepackage{hyperref}

\usepackage{graphicx}  
\usepackage{float}
\usepackage{color}

\usepackage{amssymb}

\newcounter{bla}

\begin{document}

\begin{frontmatter}



\title{Integration of Machine Learning-Based Plasma Acceleration Simulations into Geant4: A Case Study with the PALLAS Experiment}


\author[a]{A. Sytov\corref{author}}
\author[b]{K. Cassou}
\author[b]{V. Kubytskyi}
\author[b]{M. Lenivenko}
\author[c]{A. Huber}

\cortext[author] {Corresponding author.\\\textit{E-mail address:} alexei.sytov@cern.ch}
\address[a]{INFN Ferrara Division, Via Saragat 1, Ferrara, 44122, Italy}
\address[b]{Laboratoire de Physique des 2 Infinis Irène Joliot-Curie - IJCLab - UMR9012 - Bât. 100 - 15 rue Georges Clémenceau 91405 Orsay cedex - France}
\address[c]{Université de Bordeaux, CNRS, LP2I Bordeaux, 33170 Gradignan, France}

\begin{abstract}

We present the development and integration of a Machine Learning (ML)-based surrogate model, trained on Particle-In-Cell (PIC) simulations of laser-driven plasma wakefield acceleration source of electrons, into Geant4 simulation toolkit. Our model enables the generation and tracking of plasma-accelerated beams within complete experimental setups, unifying plasma acceleration and Monte Carlo-based simulations, which significantly reduces their complexity and computational cost.

Our implementation focuses on the PALLAS laser-plasma accelerator test facility, integrating its full experimental setup into Geant4. We describe the ML model, its integration into Geant4, and key simulation results, demonstrating the feasibility of start-to-end simulations of plasma acceleration facilities and applications within a unified framework.
\\

   \end{abstract}
\end{frontmatter}

\section{Introduction}
\label{intro}

Laser-driven plasma wakefield acceleration (LWFA) \cite{Tajima} is a revolutionary technology, with the potential to miniaturize particle accelerators and significantly reduce costs. The development of compact and ultrafast electron sources in the energy range of $150–250\,$MeV is of growing interest for synchrotron light sources, free-electron lasers, very high-energy electron (VHEE) radiotherapy, and secondary radiation generation [REFS]. For instance, the development of LWFA-based radiation sources is the main goal of EuPRAXIA project \cite{EuPRAXIA}, DESY \cite{DESY}, PALLAS \cite{PALLAS,drobniak2023fast}.

Coupling a laser-plasma source to practical applications necessitates precise beam transport and shaping, particularly for applications such as free-electron lasers (FELs) or ring injection. High-energy electron sources intended for direct use must ensure compact design and stable beam parameters, which are critical for applications in radiation generation and medical technologies. Achieving these goals demands comprehensive simulations that accurately model entire systems and their parameters.

The nonlinear, multi-parametric, coupled physical processes involved in laser wakefield acceleration necessitate using particle-in-cell (PIC) codes for simulating and optimizing electron sources. These codes require substantial computational resources, limiting comprehensive start-to-end studies. Recent advances in physical approximations and low-fidelity PIC simulations have enabled the generation of datasets for training highly efficient surrogate Machine Learning (ML) models. It has been demonstrated that a relatively small dataset ($\sim$ 500 configurations) is sufficient for high-performance models \cite{drobniak2023fast,kane2024surrogate}. These surrogate models can be extended to various beam parameters and serve as source terms for beam transport or interaction and tracking codes.

Geant4 \cite{Geant41,Geant42} is a comprehensive C++ Monte Carlo simulation toolkit for modeling particle interactions with matter supporting numerous applications in accelerator, nuclear, medical, high-energy physics and space science. It allows one to design and to simulate complex experimental setups including particle trajectories, electromagnetic fields, and interactions with accelerator line components such as magnets, collimators, and detectors. However, by now Geant4 lacks built-in plasma acceleration physics.

This limitation arises from the contrast between PIC methods, which solve beam-field interactions using partial differential equations, and Monte Carlo methods, which treat particles independently without collective beam effects. It is possible to merry these methods efficiently by integrating plasma acceleration into Geant4 as a particle source using ML models trained on PIC datasets. Consequently ML learning models will be able to generate charged particles of plasma accelerated beam according to the distribution arising from the LWFA system parameters.

In this paper, we demonstrate the integration of a surrogate machine learning (ML) model of a laser wakefield acceleration (LWFA) electron source into Geant4, enabling start-to-end simulations of LWFA-based applications. Specifically, we train a multilayer perceptron neural network using particle-in-cell (PIC) simulation data from the LWFA electron source in the PALLAS experiment. This trained model is incorporated into Geant4 via the Geant4 Particle Gun, allowing the neural network to generate electrons based on the parameters of the LWFA system. Additionally, we implement the full beam transport and detector system, conducting simulations of the entire experimental setup.

\section{PALLAS experimental setup}
\label{pallas}

The PALLAS project aims to develop technological building blocks for a 10 Hz laser-plasma injector electron source to achieve control and reliability comparable to RF accelerators. These developments focus on controlling the laser driver, structured plasma targets for high average power, and improving beam quality through advanced electron beam characterization.

The PALLAS electron source leverages controlled and truncated ionization injection within the plasma wakefield generated by the ponderomotive force of the laser pulse. This injection technique is widely used to control the injection volume and efficiency. The prototype compact laser-plasma electron source, installed at IJCLab, is driven by a $50\,$TW chirped pulse amplification laser system producing $2\,$J, $35\,$fs pulses at a $10\,$Hz repetition rate. The stretched pulse is amplified to approximately $3\,$J and transported via a vacuum optical transport line to a dedicated two-grating compressor in the radiation-shielded experimental hall. Spatial phase distortions of the compressed pulse are corrected using a large-aperture, high-dynamic-range deformable mirror.
 
An ultra-fast large-aperture laser pointing stabilisation system based on ultra-thin and lightweight SiC mirrors stabilizes the focal spot within a few microradians. A focusing chamber coupled with a laser diagnostic station starts the PALLAS laser-plasma injector beamline. Diagnostics in the station enable single-shot measurements of key laser pulse parameters, including integrated energy, near- and far-field spatial profiles, spatial phase, temporal profile, spectral phase, and pulse front tilt.

The laser pulse is focused by an off-axis parabola with an f/23 aperture onto the second element of the beamline, the plasma target module. The plasma module features a differential pumping system that allows continuous light gas injection, achieving interaction zone gas pressures of up to $120\,$mbar. Targets include multi-cell structured cells or gas channel types, enabling fine profiling of plasma density along the longitudinal axis. The beamline is designed to accommodate various target types directly integrated into the beamline, minimizing its footprint and allowing the placement of a standard electromagnetic quadrupole magnet close to the target.

The beam capture section includes three quadrupole magnets for variable beam focusing between $2.5$ and $4.5\,$m from the source. A fourth quadrupole is used to symmetrize the transverse beam profile. The beamline features standard electron beam diagnostics, including two beam position monitors (BPMs), an integrating current transformer (ICT), three multi-screen stations, and a resistive dipole spectrometer located $3\,$m from the source. The beamline ends with a dump designed for continuous $10\,$Hz operation with beams up to $300\,$MeV and $100\,$pC.

The machine’s control and acquisition system is based on the Tango distributed control and command system, enabling 10 Hz control and acquisition of all machine components.

Developing compact, ultrafast electron sources in the $150–250\,$MeV energy range for emerging applications is the core of the PALLAS project. Laser-plasma technology becomes competitive with RF accelerators in the energy range of $100-500\,$MeV. Stabilizing beam energy and brightness can be achieved by leveraging the chromaticity of the capture section and a collimation system. This simple approach is under technical design in the PALLAS beamline.  

Comprehensive simulations of the entire PALLAS beamline, including laser and plasma parameters, beam transport, and detection, are essential for predicting experimental results and optimizing performance.

\section{Neural network model}
\label{nn}

Our idea of the integration of ML-based plasma acceleration simulations into Geant4 requires the following steps:

\begin{itemize}
  \item To perform the PIC simulations to generate the dataset containing the output beam distribution depending on the parameters of laser-plasma system;
  \item To train the neural network using this dataset;
  \item To implement this neural network into Geant4 as a particle source with the possibility to use the parameters of the laser and plasma as an input;
  \item To exploit this particle source in specific applications of laser-plasma acceleration as Geant4 examples.
\end{itemize}

\subsection{Neural network training}

To build a neural network, two datasets of Particle-in-Cell (PIC) simulations were used: SET1 \cite{drobniak2023fast} of 12004 simulations and SET2 \cite{kane2024surrogate} of 3536 simulations. Focal position $x\_of[\mu m]$, normalized vector potential $a\_0$, dopant percentage $c\_N2[\%]$ and gas pressure $p\_1 [mbar]$ are simulation inputs.  From PIC simulations, we have full knowledge of electron beam, and we take these as model outputs: median energy $E\_med [MeV]$, energy spread $dE\_mad [\%]$, bunch charge $q\_end[C]$ and normalized transverse emittance in y plane $n\_emit\_y [m * rad]$. 

\begin{figure}[H]
    \centering
    \includegraphics[width=0.95\textwidth]{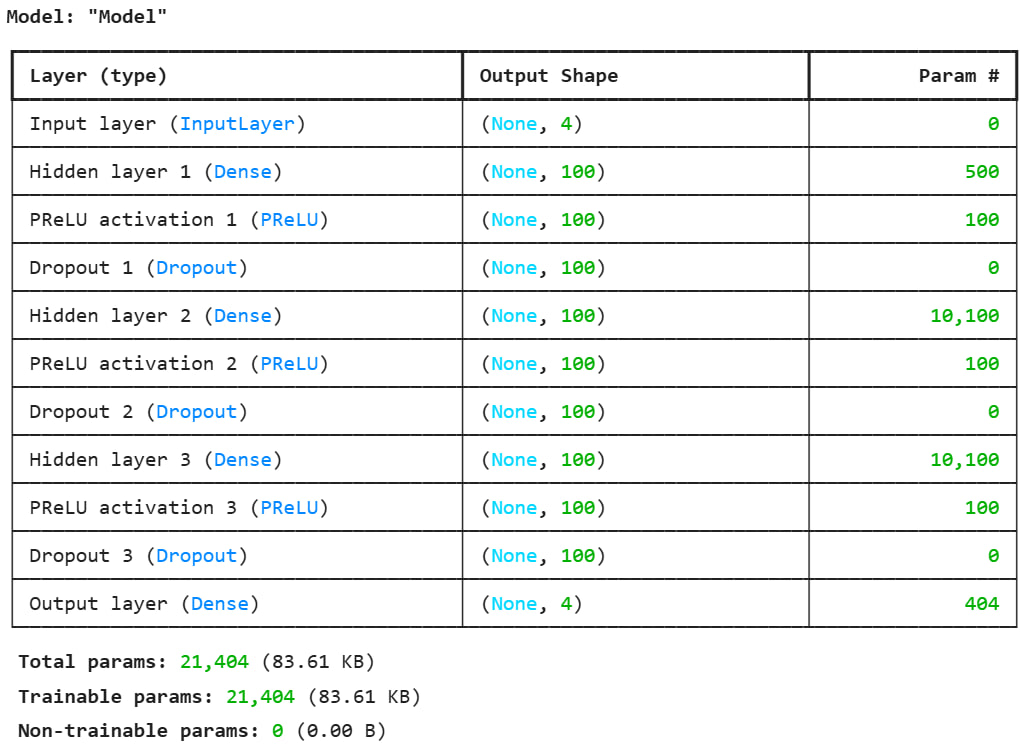}
    \caption{MLP scheme in Keras API}
    \label{fig:mlp}
\end{figure}

The multilayer perceptron (MLP) neural network was built and trained using the TensorFlow platform and Keras API\cite{tensorflow2015-whitepaper}. The MLP architecture (Fig. \ref{fig:mlp}) consists of 5 layers: an input layer with 4 neurons, 3 hidden layers with 100 neurons each, a PReLU activation function, and an output layer with 4 neurons and a sigmoid activation function.

The model was trained on the SET2 only with injection configurations. Data was scaled between 0 and 1 to ensure that all features contribute equally to the learning process and improve generalization. KFold slicing was used during training, and each slice was trained for $200$ epochs with a batch size of $50$. We used mean squared error as the loss function. The model was tested on the SET1 and achieved a coefficient of determination $98\%$.

\begin{figure}[H]
    \centering
    \includegraphics[width=0.95\textwidth]{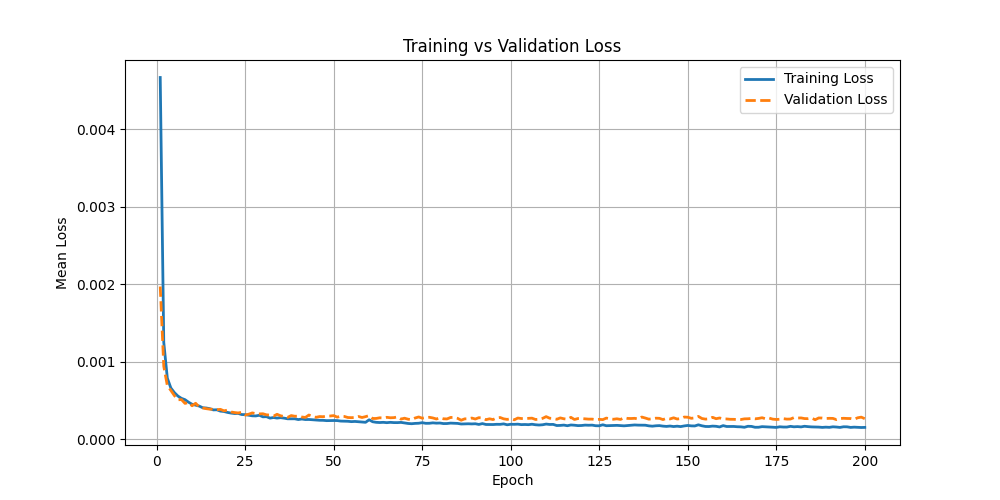}
    \caption{Mean losses for training and validation dataset for all KFolds}
    \label{fig:mlpLoss}
\end{figure}

\subsection{Implementation of the neural network into Geant4}

We used ONNX ML framework in order to implement the ML model from Python to C++, since it contains a wide library of neural network elements in both languages. In particular, our Keras model was converted into ONNX using \texttt{tf2onnx.convert.from\_keras} function in Python and saved into a file which became an input file for its equivalent into Geant4. The identity of outputs of both ML models for the same input was verified. The ONNX model was saved into a file which became an input file for model equivalent into Geant4.

In our example, the Geant4 ONNX model was implemented into Geant4 Particle Gun with the same inputs and outputs as described above. The inputs are controllable from Geant4 macro through the following commands with the standard parameters. These inputs include the option of the ML model activation and mentioned above laser focus offset, normalized vector potential, dopant percentage and gas pressure:
\\
\texttt{/gun/SetStatusONNX true \# to activate the ONNX ML model}
\\
\texttt{/laser/SetOffsetLaserFocus 558 \# um}
\\
\texttt{/laser/SetNormVecPotential 1.43}
\\
\texttt{/laser/SetFracDopTargetChamber 0.0188 }
\\
\texttt{/laser/SetPressure 58.6 \#mbar }

The ML model outputs were used in Geant4 Particle Gun to generate the energy and the transverse beam distribution.

\section{Geant4 pallas example}
\label{Geant4}

\subsection{PALLAS geometry import from STEP files to GDML files}
The geometries in Geant4 can be described using various classes that reproduce simple shapes such as parallelepipeds, spheres, cylinders, toroids, and many others, by combining these objects through Boolean operations. For more complex geometries, it is also possible to define shapes based on their boundary representation, where the solid is constructed using splines, B-splines, or tessellated surfaces.

Geant4 does not directly support import of CAD (Computer-Aided Design) geometries that fully model all elements of an experiment such as PALLAS. However, geometry files can be incorporated using a GDML (Geometry Description Markup Language) file based on XML. The GDML format consists of five sections: definition, materials, solids, structure, and setup \cite{GDML,GUIMesh}. The geometry defined in this manner can then be imported into Geant4 using the GDMLParser function.

There are various tools available to convert STEP files \cite{STEP}, derived from CAD geometries, into GDML files, such as FASTRAD \cite{FASTRAD}, ESABASE2 \cite{ESABASE}, or STEP-Tools \cite{Step-tools}. However, these tools are neither open-source nor free. For this reason, in this work, we used FreeCAD \cite{FreeCAD}, an open-source CAD editor with a user-friendly graphical interface. FreeCAD allows for the meshing of geometries imported from STEP files. Several meshing algorithms are available in the software. In our case, we used the standard algorithm, which offers two degrees of freedom: maximum surface deviation (1 mm by default) and maximum angular deviation (30° by default). These parameters were adjusted depending on the required precision or the need to limit file size, based on the specific geometric elements considered. Indeed, the finer the desired precision, the greater the number of triangles used to mesh the geometry, resulting in a larger GDML output file.

\begin{figure}[H]
\centering
\includegraphics[width=0.9\textwidth]{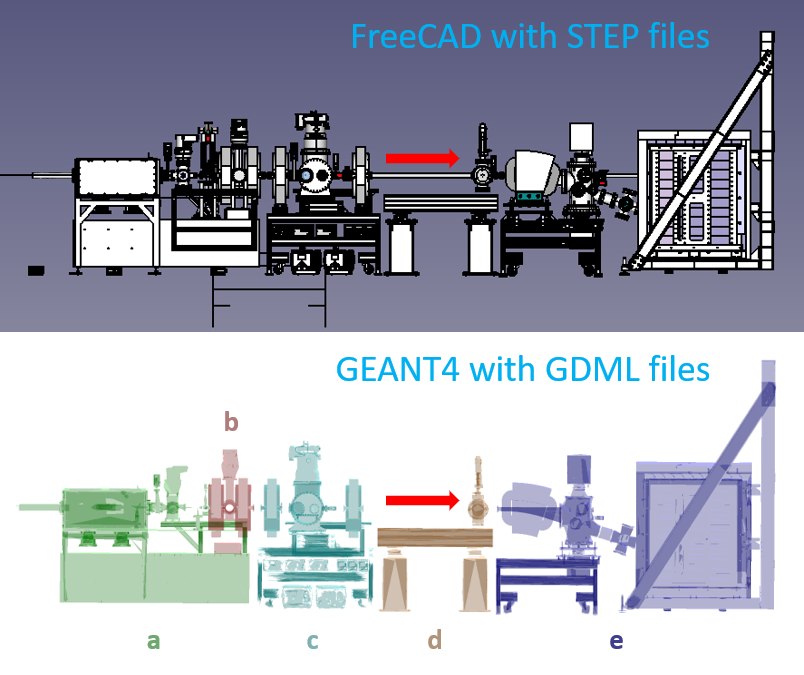}
\caption{Visualization of the PALLAS line via FreeCAD (top) and Geant4 (bottom). In Geant4 setup each color block corresponds to a geometry block that can be activated or not: (a) LIF (Laser Injection Focalisation) module, where the electron beam is produced and accelerated; (b) first focusing quadrupole pair Q1 \& Q2; (c) second focusing quadrupole pair (quadrupoles Q3 \& Q4) with the chamber where the laser is removed; (d) free space available for diagnostics detectors (Induction Current Transformer, Beam Position Monitoring and YAGs screens); (e) Dipole Chamber (including the dipole electron spectrometer with YAGs detectors) and the "dump" including shielding and chassis.}
\label{fig:PALLAS_G4}
\end{figure}

The entire PALLAS beamline was successfully loaded into Geant4 (see Figure~\ref{fig:PALLAS_G4}). Errors in particle tracking within the geometries can occur in Geant4 due to particles becoming trapped near triangular faces. In Geant4, a particle is considered trapped if its position remains unchanged for 10 consecutive steps. When this situation arises, the simulation displaces the particle by 10$^{-7}$ mm to determine whether it becomes unstuck, which was the case in our simulations.

\subsection{Implementations}

The simulation is initiated via a macro (an ASCII file) containing user interface commands, enabling easier control of the simulation and associated parameters without requiring interaction with the program's source code. Consequently, the code is designed to be accessible to users with basic knowledge of running a program, without necessarily requiring expertise in object-oriented programming or C++.

The program can be run in either sequential or multithreading mode, which can be particularly necessary for computationally expensive simulations. In computing centers equipped with machines featuring numerous cores or CPUs, the time savings increase significantly with the use of multithreading.

The simulation can also be run in two different modes. The first is a Geant4 interactive mode with visualization, while the second is a batch mode without visualization, used for data production.

The entire geometry of the PALLAS beamline was incorporated into the source code via GDML files. It is then up to the user to activate or deactivate specific elements depending on the simulation’s objective. Indeed, it may not be necessary to load the memory with the initial elements of the beamline if the study focuses on its final components. To achieve this, the user can disable the loading of certain geometry elements using specific commands in the macro.

The simulation example uses the reference physics list QGSP-BIC-HP \cite{physicsG4}. This list includes the standard physical processes for electromagnetic and hadronic physics and is specifically designed for neutron transport, covering energies from below 20 MeV to thermal energies. Thus, all processes such as ionization, Coulomb scattering, Bremsstrahlung radiation, Compton scattering, the photoelectric effect, pair production, annihilation, radioactive decay, radiative capture, fission, and elastic and inelastic hadronic scattering are taken into account in the simulation. Additionally, it may be considered to add optical processes in the future to study the response of the YAG detectors present on the PALLAS beamline in more detail.

Regarding particle generation, the code provides different use cases that can be selected by the user via the macro. Indeed, in addition to the direct use of a ParticleGun with parameters derived from the Machine Learning model, the simulation also offers particle generation through GPS (GeneralParticleSource) via macro commands. The simulation also allows particle generation from an input file, which was the method used prior to the implementation of the ONNX model directly in Geant4.

To extract information from the simulation, we used the ROOT software due to its ease of integration with Geant4. The simulation collects various data during its execution, which are then organized into different NTUPLES for future analysis.

\begin{figure}[H]
    \centering
    \includegraphics[width=0.95\textwidth]{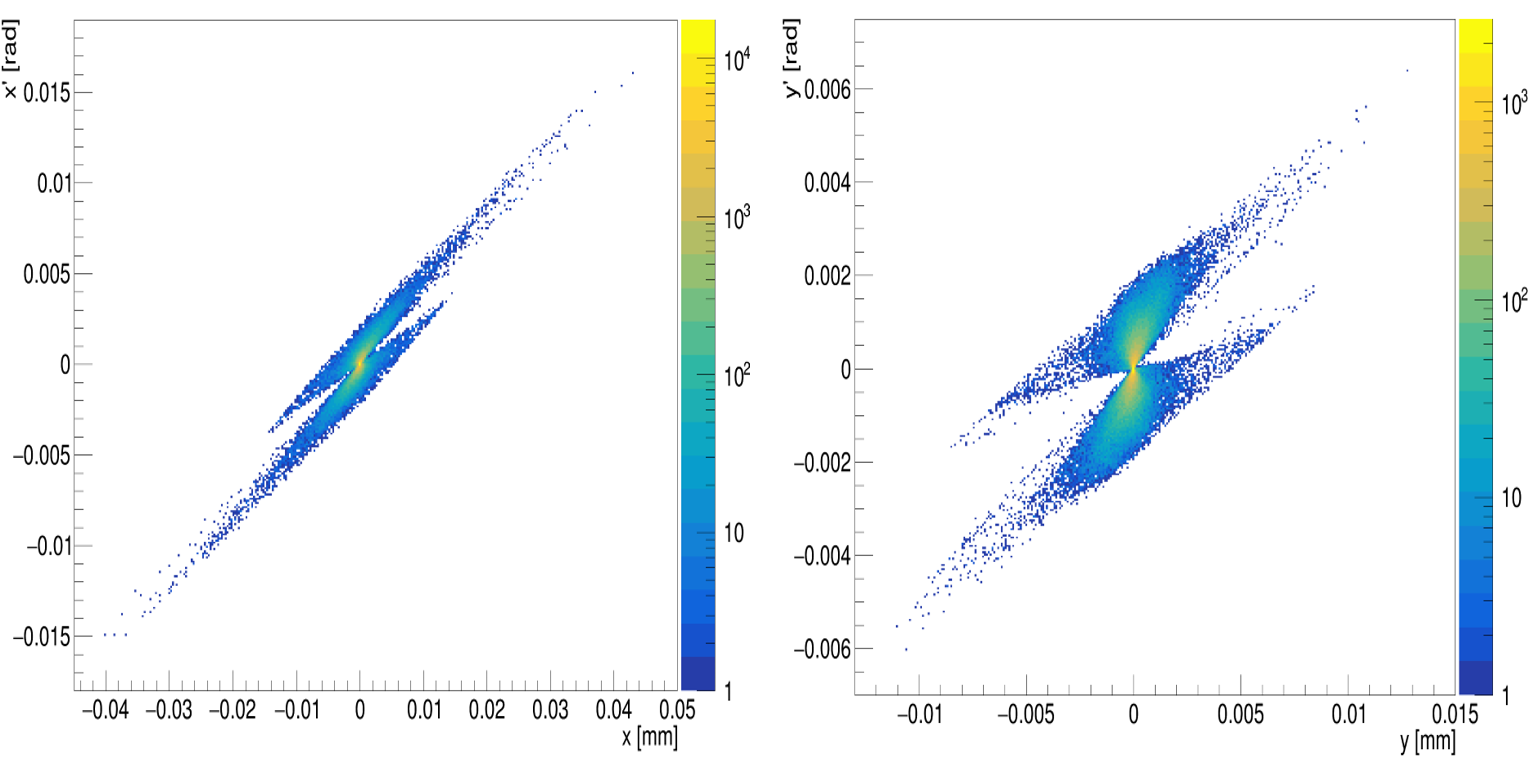}
    \caption{Example of phase spaces obtained for electron generation for a given dataset.}
    \label{fig:phasespace}
\end{figure}

\begin{figure}[H]
    \centering
    \includegraphics[width=0.8\linewidth]{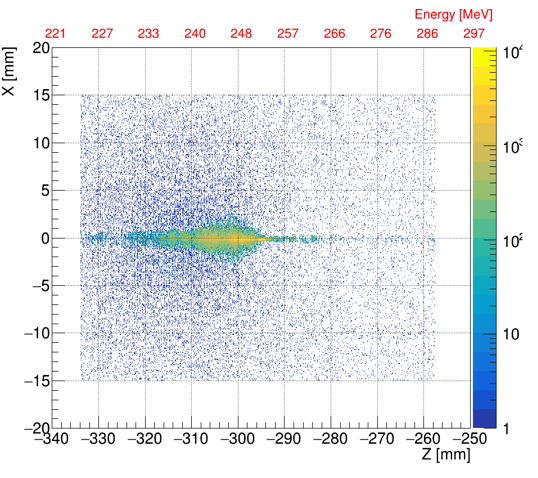}
    \caption{Example of results concerning the position of particle interaction with the YAG detector.}
    \label{fig:YAG}
\end{figure}

The simulation code provides the outputs that characterize the beam quality, a critical metric for evaluating the performance of any plasma acceleration system. Among these outputs are the phase space distributions of particles generated by our ML model (Figure~\ref{fig:phasespace}) and the spatial distribution of particles recorded on a YAG detector positioned downstream of a dipole spectrometer (Figure~\ref{fig:YAG}). The phase space distribution offers valuable insights into the transverse emittance of the beam, a key parameter for assessing beam quality. Meanwhile, the YAG detector output simulates an experimental measurement, capturing both the transverse and longitudinal (energy) distributions of the beam. This dual-output approach enables a comprehensive analysis of beam dynamics, bridging the gap between simulation and a real experimental case.

\section{Conclusion and discussion}
\label{conclusion}

Our work demonstrates successful integration of a machine learning-based surrogate model, trained on Particle-In-Cell (PIC) simulations of laser-driven plasma wakefield acceleration, into the Geant4 simulation toolkit. Our implementation, centered around the PALLAS laser-plasma accelerator facility, enables comprehensive start-to-end simulations, unifying plasma acceleration physics with Monte Carlo-based modeling.

The surrogate model, a multilayer perceptron neural network, was trained on a dataset of PIC simulations of laser-plasma interaction. ONNX framework ensured seamless translation of the ML model from Python to C++, facilitating its integration into Geant4's Particle Gun. GDML files were exploited to implement an entire PALLAS experimental setup in Geant4, including a full beam transport and detector system. 

This integration reduces the complexity and computational cost typically associated with high-fidelity plasma acceleration simulations, providing a scalable and adaptable framework for future experiments and applications. Our work paves the way for more accessible and reliable modeling of plasma-based accelerators, with potential applications in synchrotron light sources, free electron lasers, nuclear physics, and advanced radiotherapy. 





\bibliographystyle{elsarticle-num}







\end{document}